\documentstyle[12pt]{article}
\begin{document}
\begin{flushright}
DPKU9205 \\ BUTP-92/08  \\
March, 1992
\end{flushright}
\vspace{ 2cm}
\begin{center}
{\LARGE\bf  Strong CP-Problem in Superstring Theory}\footnote{ This work
was partially supported by Swiss National Science Foundation
and Japan Society for the Promotion of Science.}  \\
\vspace{2 cm}
{\Large Daijiro Suematsu}\\
\vspace {1cm}
    Institute for Theoretical Physics, University of Bern,\\
     Sidlerstrasse 5, CH-3012 Bern, Switzerland\footnote{Present address:
Department of Physics, Kanazawa University,
        Kanazawa 920, Japan.  }  \\
\end{center}
\vspace{4cm}
{\Large\bf Abstract}\\
We apply the  solution for
the strong CP-problem in the 4-dimensional superstring theory recently proposed
by Ib${\rm\acute{a}\tilde{n}}$ez and L${\rm\ddot{u}}$st to Calabi-Yau type
models and study its phenomenological aspects.
In Calabi-Yau type models there seem to be phenomenologically difficult
problems in the axion decoupling
from the neutral gauge
currents and the compatibility between the proton stability and the
cosmological bound on the axion.
DFSZ type invisible axion mechanism which works without heavy extra colored
fields may be more promising than KSVZ axion in the viewpoint of proton
stability.

\newpage
Strong CP problem\cite{KIM} still now remains being the obscure situation
 in superstring theory.
It is not known whether PQ mechanism\cite{PEC} works or not in it.
The pseudoscalar fields which come from the antisymmetric tensor are found not
to work as the invisible axion against the initial expectation\cite{WITTEN}.
The introduction of the global $U(1)$ symmetry to the superstring is also
difficult\cite{BANK}.
Recently Ib${\rm\acute{a}\tilde{n}}$ez and L${\rm\ddot{u}}$st pointed out that
the target space modular
invariant 4-dimensional superstring theory, especially, $(0,2)$ orbifolds has
the very interesting properties for
the strong CP problem\cite{Ibanez}.
Their findings are following:
1) PQ symmetry is automatically built in the theory as the K${\rm\ddot{a}}$hler
transformation associated to the target space duality.
2) The soft supersymmetry breaking terms are real and then there is no
extra dangerous contribution to the EDM of neutron in contrast with the
ordinary supersymmetric theory.
In this note we apply their solution to Calabi-Yau type models and examine
their phenomenological aspects in detail.

The target space modular invariance\cite{KIK}  is very useful to know the low
energy
effective Lagrangian of superstring\cite{FONT}\cite{FERR}.
This invariance is known to be kept in all order of string perturbation and
is usually expected to be retained even in the non-perturbative effect.
If we define the one of K${\rm\ddot{a}}$hler moduli as $ T=R^2 +iD $
 ( $R$ is the overall radius
of the compactified manifold and $D$ is the model dependent axion
originating from the antisymmetric tensor), the
target space modular transformation is defined as
\begin{eqnarray}
  T \rightarrow T^\prime ={aT-ib \over icT+d}
\end{eqnarray}
where $ad-bc=1$ and $a,b,c,d$ are integers.\footnote{If there are
more K${\rm\ddot{a}}$hler moduli fields as the
general Calabi-Yau case, the transformation becomes more complicated one.
But (1) is expected to be contained in it.
In the following study we confine ourselves to the one K${\rm\ddot{a}}$hler
 modulus case, for simplicity.}
Under this transformation, the every chiral fields $ \phi_i $ have modular
weight $n_i$ and transform as
\begin{eqnarray}
 \phi_i \rightarrow \phi_i^\prime =(icT+d)^{n_i}\phi_i.
\end{eqnarray}
The generalized K${\rm\ddot{a}}$hler potential
${\cal G}=K(\Phi_i,\Phi_i^\ast ) +\log \vert W(\Phi_i)\vert^2 $
is known to be invariant under the following K${\rm\ddot{a}}$hler
transformation
 \begin{eqnarray}
K(\Phi_i,\Phi_i^\ast )& \rightarrow & K(\Phi_i,\Phi_i^\ast )
                         +F(\Phi_i)+F^\ast (\Phi_i^\ast ),\nonumber \\
   W(\Phi_i)& \rightarrow &\exp (-F(\Phi_i))W(\Phi_i),
 \end{eqnarray}
where $\Phi_i$ represents $T$ and matter fields $\phi_i$.
The target space modular transformation (1) induces a K${\rm\ddot{a}}$hler
transformation with $F(T)= \log (icT+d)^3$.
The superpotential $W(T,\phi_i )$ of the charged matter fields $\phi_i$ is
generally written as
\begin{eqnarray}
  W(T,\phi_i ) = \sum_{i,j,k} \lambda_{ijk}(T)\phi_i\phi_j\phi_k.
\end{eqnarray}
{}From eqs.(2) and (3), it is  required that $\lambda_{ijk}(T)$ should
 transform in the following way,
\begin{eqnarray}
 \lambda_{ijk}(T) \rightarrow (icT+d)^{-3-n_i-n_j-n_k}\lambda_{ijk}(T).
\end{eqnarray}
Here we should note that in order for the transformation (3) to be the
symmetry of the theory, the fermion fields $\psi_I$ with the canonically
normalized kinetic term should be also transformed as
\begin{eqnarray}
\psi_I \rightarrow \exp[-{1 \over 4}q_I(F-F^\ast )] \psi_I,
\end{eqnarray}
where $q_I=1$ for gauginos and $q_I=-1-{2 \over 3}n_I$ for matter fermions
($n_I$ is the modular weight)\cite{Ibanez}\cite{Derend}.
As a result, this symmetry has generally a triangle anomaly
\begin{eqnarray}
{i \over 32\pi^2}\sum_I q_I {\rm Tr}~T_I^2(F-F^\ast )F_{\mu\nu}^a
                           \tilde F^{a\mu\nu}.
\end{eqnarray}
$T_I$ is the generator of the gauge group in the representation of the
 fermion $\psi_I$.
This means that if there is a scalar field which has a flat potential
except for the effect of the color anomaly, an
invisible axion mechanism can work due to this symmetry.
In this mechanism the PQ-like symmetry is built in the effective Lagrangian of
the superstring as its own property from the beginning.

Now we apply these mechanisms to Calabi-Yau type superstring models.
In these models the low energy effective Lagrangian is
composed of the various charged matter fields.
These models have the following $E_6$ charged fields
\begin{equation}
 \begin{array}{lll}
Q(3,2)_{1/3,1/\sqrt 3,-1/\sqrt 6},
&\bar u(3^\ast ,1)_{-4/3,1/\sqrt 3,-1/\sqrt 6},
&\bar d(3^\ast ,1)_{2/3,1/\sqrt 3,3/\sqrt 6}, \nonumber  \\
\ell (1,2)_{-1,1/\sqrt 3,3/\sqrt 6},
&S_1(1,1)_{0,1/\sqrt 3,-5/\sqrt 6},
&\bar e(1.1)_{2,1/\sqrt 3,-1/\sqrt 6},  \nonumber  \\
h(1,2)_{1,-2/\sqrt 3,2/\sqrt 6},
&h^\prime (1,2)_{-1,-2/\sqrt 3,-2/\sqrt 6},
&S_2(1,1)_{0,4/\sqrt 3,0},   \nonumber  \\
g(3,1)_{-2/3,-2/\sqrt 3,2/\sqrt 6},
&\bar g(3^\ast ,1)_{2/3,-2/\sqrt 3,-2/\sqrt 6},
&
 \end{array}
\end{equation}
and also $E_6$ singlet fields.
 The above representations of each fields stand for the ones
under $SU(3)_C \times SU(2)_L\times U(1)_Y\times U(1)_\psi \times U(1)_\chi
(\subset E_6)$.\footnote{We adopt $E_6$ model as the starting point of
our following arguments.
The generalization to the models with more extra $U(1)$'s is straightforward.}
Without loss of generality the superpotential $W$ for these fields is
expressed
\begin{eqnarray}
 W &=&\lambda_1 QQg +\lambda_2 Q\bar uh +\lambda_3 Q\bar dh^\prime
   +\lambda_4 Q \bar g\ell
   +\lambda_5 \bar u\bar g \bar d +\lambda_6 \bar ug\bar e \nonumber \\
 & &  +\lambda_7\bar dgS_1 +\lambda_8\bar ggS_2
   +\lambda_9 hh^\prime S_2 +\lambda_{10} \ell hS_1 +
    \lambda_{11}\ell h^\prime \bar e.
\end{eqnarray}
We abbreviated the generation indices.
$\lambda_2, \lambda_3, \lambda_7$ and $\lambda_8$ terms are relevant to
$\bar \theta =\theta + arg~det M$ through the mass matrix $M$ of the
colored fields.
As seen from eqs.(5) and (7), under the target space modular transformation
(1)  parameter $\theta$ shifts
through these couplings
 \begin{eqnarray}
  \theta &\rightarrow& \theta
  + \log det({\lambda_8 (T)S_2 \over \lambda^\ast_8 (T)S_2^\ast })
  + \log det({\lambda_2 (T)h \over \lambda^\ast_2 (T)h^\ast })
  + \log det({\lambda_3 (T)h^\prime \over \lambda^\ast_3 (T)
                            h^{\prime\ast} })
   \nonumber    \\
& &=\theta +\sum_{j,k}(-3-n_j-n_k)
                      \log ({icT+d \over -icT^\ast +d}).
  \end{eqnarray}
The summation should be taken over the colored fields contributing to these
couplings.
To realize the symmetry breaking of the standard model in our
considering models, all of $S_1, S_2, h, h^\prime$ must have VEVs.
This suggests that the linear combination of the phase of $S_1, S_2, h$ and
$h^\prime$ will work as an axion.\footnote{In general $\lambda_7$ term is
phenomenologically unfavorable and we assume
it to be zero in eq.(10) and the following arguments.
Particurally, $S_2$ should have a VEV at an intermediate mass scale $\sim
10^{11}$GeV.}

In the most superstring theories it is known that there exist the extra color
triplets $g, \bar g$
which couple to the singlet fields similar to $S_2$ as $\lambda_8$ term.
Ib${\rm\acute{a}\tilde{n}}$ez and L${\rm\ddot{u}}$st pointed out that the phase
part of the K${\rm\ddot{a}}$hler transformation associated to the target space
modular
transformation (1)
plays the $U(1)_{PQ}$ role through $\bar ggS_2$ coupling like the
invisible axion model of KSVZ\cite{KIMA}.
As seen from the above arguments we should note that the axion in this
mechanism has the following properties:
\begin{description}
\item{(i)} ordinary quarks, leptons and doublet Higgs scalars have this
$U(1)_{PQ}$
charge as DFSZ axion model\cite{DINE},
\item{(ii)} singlet field $S_2$ decouples from ordinary quarks
and leptons not due to $U(1)_{PQ}$ but the extra gauge symmetry,
\item{(iii)} extra triplets $g$ and $\bar g$ generally couple not only to
singlet
$S_2$ but also to ordinary quarks
and leptons through $\lambda_1,\lambda_4,\lambda_5$ and $\lambda_6$ terms.
\end{description}
These properties yield the non-trivial phenomenological problems on this
mechanism at least in the Calabi-Yau type models.
That is, (i) makes the axion the mixture of doublet Higgs $h,h^\prime $ and
singlets $S_1,S_2$.
As a result there appears non-trivial axion decoupling problem from the
neutral gauge currents in Calabi-Yau type models.
Here we should note that the extra gauge symmetry also plays an important role
 to guarantee the flatness of the scalar potential of $S_2$ ({\it i.e.} the
absence of
$S_2^3$ term in $W$) other than (ii).
(iii) brings the compatibility problem of the proton stability and the
cosmological bound on the axion.
In the following, we study these problems in detail.

We  start from the study of the axion decoupling from the neutral gauge
currents in this scenario.
At first, we briefly review the generalized axion model\cite{KIM}.
Let's consider the models which have a set of scalar fields $\phi_i$ with
non-trivial PQ charge $\Gamma_i$. We define
\begin{eqnarray}
 \phi_i={1 \over \sqrt 2}(v_i +\eta_i)\exp (i{\xi_i \over v_i})
\end{eqnarray}
where $v_i$ is the vacuum expectation value(VEV).
The axion field is written as
\begin{eqnarray}
 a={1 \over f_a}\sum_i \Gamma_i v_i\xi_i .
\end{eqnarray}
 $f_a$ is the axion decay constant and expressed by
\begin{eqnarray}
 f_a^2 =\sum_i \Gamma_i^2v_i^2.
\end{eqnarray}
We should note that $f_a$ is dependent on the absolute value of the PQ
charge.
Only if $\Gamma_i =O(1)$, $f_a$ is the measure of the scale of symmetry
breaking.
When the symmetry breakings occur, would-be Nambu-Goldstone bosons are absorbed
by the gauge bosons.
The axion must be orthogonal to these neutral gauge currents $j_\mu^\alpha $,
\begin{eqnarray}
 \langle 0 \mid j^\alpha_\mu \mid a \rangle =0.
\end{eqnarray}

Now we apply these to our study.
Let's the PQ charges of the relevant scalar fields $h,h^\prime,S_1,S_2$ be
$\Gamma^{(h)},\Gamma^{(h^\prime)},\Gamma^{(1)},\Gamma^{(2)}$, respectively.
The PQ current is
 \begin{eqnarray}
j^a_\mu &=& i(\sum_i\Gamma^{(h)}_ih_i^\dagger\stackrel{\leftrightarrow}
                                             \partial_\mu h_i
+\sum_j\Gamma^{(h^\prime )}_jh_j^{\prime \dagger}\stackrel{\leftrightarrow}
                                             \partial_\mu h_j^\prime
+\sum_k\Gamma^{(1)}_kS_{1k}^\dagger\stackrel{\leftrightarrow}
                                   \partial_\mu S_{1k}
 +\sum_l\Gamma^{(2)}_lS_{2l}^\dagger\stackrel{\leftrightarrow}
                                     \partial_\mu S_{2l}) \nonumber  \\
  & &+{\rm (extra\  singlet\  and\  quark/lepton\  currents)}   \nonumber \\
   &=&\sum_i\Gamma^{(h)}_iv^{(u)}_i\partial_\mu\xi_i^{(u)}
   +\sum_j\Gamma^{(h^\prime )}_jv^{(d)}_j\partial_\mu\xi_j^{(d)}
    +\sum_k\Gamma^{(1)}_ku^{(1)}_k\partial_\mu\xi^{(1)}_k
    +\sum_l\Gamma^{(2)}_lu^{(2)}_l\partial_\mu\xi^{(2)}_l  \nonumber \\
  & &+{\rm (extra\  singlet\  and\  quark/lepton\  currents)}   \nonumber   \\
  &=&f_a\partial_\mu a
   +{\rm (extra\  singlet\  and\  quark/lepton\  currents)},
  \end{eqnarray}
where $i,j,k,l$ are the generation indices.
$v^{(u)}, v^{(d)}, u^{(1)} $ and $u^{(2)}$ are the VEVs of $h, h^\prime , S_1$
and $S_2$, respectively.
If there are some extra singlets, those contributions should be included as
indicated in the parentheses.
The axion field is
\begin{eqnarray}
 a={1 \over f_a}
    (\sum_i\Gamma^{(h)}_iv^{(u)}_i\xi_i^{(u)}
+\sum_j\Gamma^{(h^\prime )}_jv^{(d)}_j\xi_j^{(d)}
+\sum_k\Gamma^{(1)}_ku^{(1)}_k\xi^{(1)}_k
+\sum_l\Gamma^{(2)}_lu^{(2)}_l\xi^{(2)}_l),
\end{eqnarray}
where
\begin{eqnarray}
 f_a^2=\sum_i(\Gamma^{(h)}_iv^{(u)}_i)^2
+\sum_j(\Gamma^{(h^\prime )}_jv^{(d)}_j)^2
+\sum_k(\Gamma^{(1)}_ku^{(1)}_k)^2
+\sum_l(\Gamma^{(2)}_lu^{(2)}_l)^2.
\end{eqnarray}
We can generally consider the gauge structure
$SU(3)_C\times SU(2)_L\times U(1)_Y\times U(1)^2(\subset E_6)$ and also take
the
$\chi ,\psi $ basis with respect to extra $U(1)^2$ as defined in
(8).\footnote{It should be noted that there are necessarily two extra $U(1)$
factors in the models with the intermediate mass scale\cite{MATSU} whose
existence is
the necessary condition for our mechanism.}
There are three neutral currents $j^{Z^0}_\mu , j^\chi_\mu ,
j^\psi_\mu$ which couple to $Z^0$ and extra $U(1)^2$ gauge bosons,
respectively.
The axion should decouple from these currents,
\begin{eqnarray}
 \langle 0\mid j^{Z^0}_\mu \mid a\rangle =
   \langle 0\mid j^\chi_\mu \mid a\rangle =
\langle 0\mid j^\psi_\mu \mid a\rangle =0.
\end{eqnarray}
The axion should also be orthogonal to $\pi_0$. But here we donot discuss
this condition.
If there are more extra gauges, the additional decoupling conditions must
be imposed on the axion field $a$.
The neutral currents are expressed as
 \begin{eqnarray}
j^{Z^0}_\mu&=& {1 \over v}(\sum_iv^{(u)}_i\partial_\mu \xi^{(u)}_i
                -\sum_jv^{(d)}_j\partial_\mu \xi^{(d)}_j), \nonumber \\
j^{\chi}_\mu&=&{1 \over u}
   ({2 \over \sqrt 6}\sum_iv^{(u)}_i\partial_\mu \xi^{(u)}_i
   -{2 \over \sqrt 6}\sum_jv^{(d)}_j\partial_\mu \xi^{(d)}_j
   -{5 \over \sqrt 6}\sum_ku^{(1)}_k\partial_\mu \xi^{(1)}_k),\\ \nonumber
j^{\psi}_\mu &=&{1 \over w}
    ({-2 \over \sqrt 3}\sum_iv^{(u)}_i\partial_\mu \xi^{(u)}_i
      -{2 \over \sqrt 3}\sum_jv^{(d)}_j\partial_\mu \xi^{(d)}_j
    +{1 \over \sqrt 3}\sum_ku^{(1)}_k\partial_\mu \xi^{(1)}_k
   +{4 \over \sqrt 3}\sum_lu^{(2)}_l\partial_\mu \xi^{(2)}_l).
   \end{eqnarray}
$u, v$ and $w$ are defined in the similar way as eq.(13).
Substituting eqs.(16) and (19) into eq.(18),
we get the decoupling conditions
 \begin{eqnarray}
\sum_i\Gamma^{(h)}_iv^{(u)2}_i-\sum_j\Gamma^{(h^\prime )}_jv^{(d)2}_j
       & =&0,  \nonumber  \\
   \sum_k\Gamma^{(1)}_ku^{(1)2}_k&=&0,  \\ \nonumber
   \sum_i\Gamma^{(h)}_iv^{(u)2}_i- \sum_l\Gamma^{(2)}_lu^{(2)2}_l&=&0.
  \end{eqnarray}
In our scenario the PQ symmetry is built in the theory from the beginning.
Therefore there is no freedom to tune the PQ charge so as to guarantee the
existence of the axion decoupling from the neutral gauge currents correctly.
Depending on the model it is determined automatically whether the invisible
axion exists or not.

To study this problem, we need to determine the PQ charge of each scalar
fields $\phi_i$.
{}From the fact that the phase part of eq.(2) corresponds to $U(1)_{PQ}$, PQ
charge of $\phi_i$ is
\begin{eqnarray}
  \Gamma_i =n_i\tan^{-1}({T_R \over \alpha -T_I}).
\end{eqnarray}
$\alpha$ is an arbitrary parameter and $T_R, T_I$ are real and imaginary
parts of $T$, respectively.
$\tan^{-1}({T_R \over \alpha -T_I})$ is considered as the normalization factor
of the
PQ charge.
 For the untwisted
matter fields $n_i$ equals to $-1$ and the twisted matter fields have
$n_i \leq -1$ integer values depending on
the way of the twists.
 Calabi-Yau type models have no massless twisted matter fields.
Therefore the modular weights have the same sign and then the PQ charges have
also the same sign for all matter fields.
As a result the axion decoupling condition (20)
cannot be satisfied realizing the non-trivial hierarchy
\begin{eqnarray}
 v^{(u)} \sim v^{(d)} < u^{(1)} \ll u^{(2)}.
\end{eqnarray}
This hierarchical structure is necessary to bring the symmetry breaking
pattern of the standard model\footnote{We should note that $S_1$ cannot have a
VEV at high energy scale like $S_2$ because $S_1$ has no D-term flat potential.
See also the arguments in the next paragragh.}\cite{MATSU}.
Thus it is difficult for our axion mechanism to work in the
Calabi-Yau type models.

Usually in the 4-dimentional superstring there exist extra $U(1)$ factors and
new matter singlets other than those of the $E_6$ models and we can
expect thses ingredients to remedy the situation discussed above in the
Calabi-Yau case.
In that case the restriction on the superpotential due to the discrete symmetry
will be necessary.
For example, the role of extra gauge symmetry in (ii) should be played by the
discrete symmetry.

Next we study the compatibility of the proton stability and the axion
cosmological bound.
As shown in eq.(9) , $g$ and $\bar g$ couple to the ordinary quarks and
leptons. And
these couplings induce the proton decay through the tree level couplings
$O({\lambda_1\lambda_4 \over M_g^2})(QQQ\ell )$ and/or
$O({\lambda_5\lambda_6 \over M_g^2})(\bar u\bar u\bar d\bar
e)$\cite{MATSU}\cite{DINEA}.
Since Yukawa coupling constants $\lambda_i$ are usually order one\cite{DIS}
, these amplitudes are determined by $M_g$ which is estimated through $S_2\bar
gg$ coupling as $M_g=\lambda_8\langle S_2\rangle$.
Here the intermediate mass scale $\langle S_2\rangle$ is introduced in the
following way.
The scalar potential of $S_2$ is composed of F- and D-terms.
In order to keep the supersymmetry until the weak scale, $ S_2$ must
have the
flat potential at the intermediate scale.
D-term flatness is guaranteed by $\langle S_2\rangle =\langle\bar S_2\rangle$
because the D-term
contribution to the scalar potential is
$\sum_\alpha (S_2^\dagger T_\alpha S_2 -\bar S_2^\dagger T_\alpha \bar S_2)$
where $T_\alpha$ expresses the $U(1)$ charge\cite{DINEA}.
Taking account of the absence of $S_2^3$ term in the superpotential $W$
due to the extra $U(1)$ symmetry, the scalar potential
are produced by the non-renormalizable terms\cite{Lutkin}
\begin{eqnarray}
 V= \lambda^{(p)}M_C^{6-4p}S_2^{4p-2} -M_S^2S_2^2,
\end{eqnarray}
where $M_C$ is the compactification scale and $p$ represents the lowest
order contribution.
$M_S^2S_2^2$ is the soft supersymmetry breaking term induced from the hidden
sector.
{}From this we get
\begin{eqnarray}
 \langle S_2\rangle =M_C({M_S \over M_C})^{1 \over (2p-2)}.
\end{eqnarray}
The present experimental bound of the proton stability requires $\langle
S_2\rangle
>10^{16}$GeV.
This is realized only when $p \geq 4$.
And such a condition is known to be satisfied in some models which have the
large symmetry.
The cosmological bound on the axion model imposes on the
axion decay constant $f_a<10^{12}$GeV\cite{PRES}.
If $f_a \sim \langle S_2\rangle$ as expected from eq.(13), the discrepancy
occurs between the proton
stability and the axion cosmological bound.

In order to make our axion mechanism realistic we must reconcile these.
The most simple solution for this will be the following one.
The tree level amplitude of the proton decay is zero because the relevant
Yukawa couplings in the superpotential $W$ are zero.
And $f_a \sim \langle S_2\rangle \sim \sqrt{M_CM_S} \sim 10^{11}$GeV ($p=2$).
This possibility has been studied for the proton stability
in the various superstring models.
But within our knowledge the realistic model which realizes this has not
been found by now.

We may consider the other type solution for this problem.
As mentioned before, the axion decay constant $f_a$ is not the measure of the
symmetry breaking scale.
They are related by eq.(13) and in our case PQ charge is given by eq.(21).
{}From the study of the effective theory with target space modular invariance
\cite{FONT},  we know $\langle T_I\rangle =0, \langle T_R\rangle =O(1)$ and
then
$ \Gamma_i \sim n_i \tan^{-1}(1/\alpha )$.
If this PQ charge normalization is taken to be extremely small({\it i.e.}
$\alpha \sim 10^{5}$) , we can realize
simultaneously $f_a <10^{12}$GeV and $M_g >10^{16}$GeV in the model with
$\langle S_2\rangle > 10^{16}$GeV.
However, this appears to be unnatural.
The reconsilation of these may remain potentially as a serious problem for the
present mechanism not only in Calabi-Yau type models but also other
4-dimensional superstring models unless we find the way to suppress the
coupling of the extra colored fields to ordinary quarks and leptons.

In the above consideration we assume the existence of the extra color
triplets $g$ and $\bar g$ as suggested by Ib${\rm\acute{a}\tilde{n}}$ez and
L${\rm\ddot{u}}$st.
However, following eq.(10) DFSZ type invisible axion mechanism seems to be able
to work without $g$ and $\bar g$.
If there are no $g$ and $\bar g$, the proton stability problem related to
the intermediate mass scale $\langle S_2\rangle$ will disappear and we can
take $\langle S_2\rangle \sim 10^{11}$GeV which is consistent with the
cosmological bound.
Moreover there may be a good feature that the so-called
$\mu$-problem\cite{KIMB} does not exist because there are generally two kinds
of singlet fields like $S_2$ as stressed in ref.\cite{MATSU}.
Only $S_2$ which has the partner $\bar S_2$ can have a VEV at the intermediate
scale because of the D-term flatness.
$S_2$ which has no partner $\bar S_2$ remains massless until the weak scale and
it can contribute to $\lambda_9$ term which is relevant to the symmetry
breaking at the weak scale.
In the viewpoint of proton stability this possibility may be more realistic
than KSVZ type solution in which the extra colored heavy fermions play the
important role.
Unfortunately in Calabi-Yau type models we cannot find such a solution
because $g$ and $\bar g$ are contained in {\bf 27} of $E_6$ and remain
massless at the compactification scale.
Orbifold may give such models.

In conclusion, we applyed the solution for the strong CP problem
in the 4-dimensional superstring theory recently proposed by
Ib${\rm\acute{a}\tilde{n}}$ez and L${\rm\ddot{u}}$st to Calabi-Yau type models.
Their mechanism is very interesting but in the Calabi-Yau type models
there seem to be difficult phenomenological problems.
It is very interesting to find an explict $(0,2)$ orbifold models in which
their mechanism can be realized in phenomenologically successful way.
The DFSZ type invisible axion model which works without heavy extra colored $g$
and $\bar g$ fields will be more promising than KSVZ axion
 from the viewpoint of the proton stability.
It is also interesting subject to construct such models concretely.

\newpage

\end{document}